\documentclass[prb,twocolumn,showpacs,superscriptaddress,amsmath,amssymb]{revtex4}
\usepackage{mathrsfs}

\usepackage{graphicx}
\usepackage{dcolumn}
\usepackage{bm}
\usepackage{color}

\begin{document}


\title{Induced chiral Dirac fermions in graphene by a periodically modulated magnetic field}

\author{Lei Xu}
\affiliation{National Laboratory of Solid State Microstructures and
Department of Physics, Nanjing University, Nanjing 210093, China}

\author{Jin An}
\affiliation{National Laboratory of Solid State Microstructures and
Department of Physics, Nanjing University, Nanjing 210093, China}

\author{Chang-De Gong}
\affiliation{Center for Statistical and Theoretical Condensed Matter
Physics, and Department of Physics, Zhejiang Normal University,
Jinhua 321004, China}

\affiliation{National Laboratory of Solid State Microstructures and
Department of Physics, Nanjing University, Nanjing 210093, China}

\date{\today}

\begin{abstract}
The effect of a modulated magnetic field on the electronic structure
of neutral graphene is examined in this paper. It is found that
application of a small staggered modulated magnetic field does not
destroy the Dirac-cone structure of graphene and so preserves its
4-fold zero-energy degeneracy. The original Dirac points (DPs) are
just shifted to other positions in $k$ space. By varying the
staggered field gradually, new DPs with exactly the same
electron-hole crossing energy as that of the original DPs, are
generated, and both the new and original DPs are moving
continuously. Once two DPs are shifted to the same position, they
annihilate each other and vanish. The process of generation and
evolution of these DPs with the staggered field is found to have a
very interesting patten, which is examined carefully. Generally,
there exists a corresponding branch of anisotropic massless fermions
for each pair of DPs, resulting in that each Landau level (LL) is
still 4-fold degenerate except the zeroth LL which has a robust
$4n_t$-fold degeneracy with $n_t$ the number of pairs of DPs. As a
result, the Hall conductivity $\sigma_{xy}$ shows a step of size
$4n_te^2/h$ across zero energy.

\end{abstract}

\pacs{73.43.Cd, 73.22.Pr, 73.61.Wp}

\maketitle

Low energy physics of neutral graphene is characterized by the two
inequivalent Dirac cones which is related by the time-reversal
symmetry and described by the relativistic massless Dirac
equation.\cite{Haldane1988,Zheng2002,Novoselov2005,Zhang2005} Nearly
all important properties of neutral graphene is governed by the
chiral massless fermions around the two cones. For example, the
zero-energy anomaly due to the linear energy dispersion and the
particle-hole symmetry of the Dirac cones give rise to the anomalous
quantum Hall
effect(QHE)\cite{Novoselov2005,Zhang2005,Gusynin2005,Zhou2006,Sheng2006,Novoselov2007}
or the so-called half-integer QHE, where the Hall conductivity is
quantized to be half-integer multiples of 4$e^2/h$. When the
Dirac-cone topology is destroyed or replaced by other structures,
the system will undergo quantum phase transitions. In bilayer
graphene, each Dirac cone is replaced by two touching parabolic
bands,\cite{McCann2006,Novoselov2006,Nilsson2006} which leads to the
8-fold degeneracy of zero-energy level,\cite{McCann2006} giving rise
to the quantized Hall conductivity in bilayer graphene taken on
values of integer multiples of
4$e^2/h$.\cite{McCann2006,Novoselov2006}

Modulation of electronic structure in graphene has already been
experimentally realized, where periodic
electronic\cite{Park2008,Michal Barbier2008} or
magnetic\cite{RamezaniMasir2009,DellAnna2009} potentials can be
applied to graphene by making use of
substrate\cite{Sutter2008,Vazquez2008,Martoccia2008,Pletikosic2009}
or controlled adatom deposition,\cite{Meyer} or by fabrication of
periodic patterned gate electrodes. This kind of graphene
superlattice potential can change the Dirac-cone structure of
graphene dramatically,\cite{Park2009,Brey2009} which may lead to
some new phenomena, as well as potential application of graphene
materials.

\begin{figure}
\scalebox{0.6}[0.6]{\includegraphics[105,375][451,711]{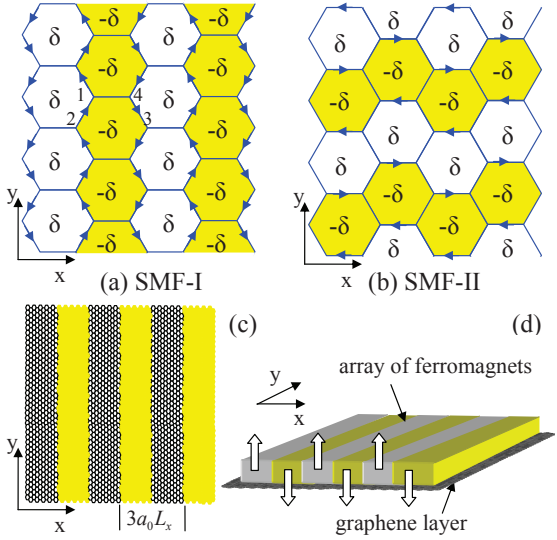}}
\caption{\label{fig1}(Color online) Illustration of the rectangular
sample of graphene under periodically modulated magnetic fields. (a)
and (b) represent two simplest SMFs, where each white and yellow
(grey) hexagon has a flux $\delta$ and $-\delta$, respectively. The
numbers $1,\cdots,4$ represent the inequivalent atoms in a unit
cell. Each arrow represents a phase shift suffered by electrons when
hopping along the direction, which is $\delta/4$ in case (a), and
$\delta/2$ in case (b). (c) represents a long-period staggered flux
applied to graphene with lattice period $3a_0L_x$, where $a_0$ is
the lattice constant. (d) is a corresponding experimental layout in
which there is an array of ferromagnetic stripes with alternative
magnetization on the top of a graphene layer.}
\end{figure}

In this paper, we present a study on the electronic structure of
monolayer neutral graphene and its unusual integer QHE under the
influence of a periodically modulated orbital magnetic field, which
is schematically shown in Fig.1. This kind of one-dimensional
modulation of magnetic field can be achieved in experiments by
applying an array of ferromagnetic stripes with alternative
magnetization on the top of a graphene layer, or by making use of
cold atoms in a honeycomb optical
lattice,\cite{Grynberg1993,Wu2007,Wu2008} or ``artificial graphene"
realized in a nanopatterned two-dimensional electron
gas.\cite{Gibertini2009} Our analysis shows that generally the
Dirac-cone structure can not be smeared out by this time-reversal
invariant magnetic field. Similar to the cases of periodic
electronic potential,\cite{Park2009,Brey2009} new DPs will be
generated with varying the amplitude of the field. What's remarkable
and different is that the newly generated DPs together with the
original DPs will move and evolve in $k$ space with the field. This
leads to a series of quantum phase transitions with each phase
characterized by its unusual integer QHE, which is expected to be
observed by Hall measurements.

We start with the tight-binding model on a honeycomb lattice in the
presence of a perpendicular, periodically modulated orbital magnetic
field. The Hamiltonian is given by,
\begin{equation}
H=-t\sum_{<ij>}e^{ia_{ij}}c_{i}^\dag c_{j}+ \texttt{H.c.},
\label{eq:one}
\end{equation}
where $c_i^\dag$ ($c_i$) is an electron creation (annihilation)
operator on site $i$, and $<ij>$ denotes nearest-neighbor pairs of
sites. Here the spin index is suppressed since we do not consider
the Zeeman splitting. The magnetic flux per hexagon (the summation
of $a_{ij}$ along the six bonds around a hexagon) is given by $\sum
a_{ij}=\phi\pm\delta$, where $\phi$ measures the uniform magnetic
flux whereas $\delta$ the staggered modulated flux, both of which
are in units of $\phi_0/2\pi$ with $\phi_0$ the flux quantum.
Hereafter energy is measured in unit of the nearest-neighbor hopping
integral $t$.

To begin with, let us consider the effect of the two simplest types
of staggered magnetic fields (SMFs), in order to extract the main
physics behind graphene under the influence of a modulated orbital
magnetic field. The configurations of the two types are
schematically shown in Fig.~\ref{fig1}(a) and (b), respectively,
where proper gauge has been chosen for each case.

We first show the evolution of DPs from an analytical calculation
for SMF-I shown in Fig.~\ref{fig1}(a). The tight-binding Hamiltonian
in $k$-space can be written as
\begin{eqnarray}
{\cal H}=\left(
\begin{array}{cccc}
0&\gamma_k&0&\eta_k\\
\gamma_k^\dag&0&\eta_k^\dag&0\\
0&\eta_k&0&\gamma_{-k}^\dag\\
\eta_k^\dag&0&\gamma_{-k}&0
\end{array}\right)
\end{eqnarray}
where
$\gamma_k=-2te^{-i\frac{k_x}{2}}\cos(\frac{\sqrt{3}}{2}k_y+\frac{\delta}{4})$
and $\eta_k=-te^{ik_x}$. The Hamiltonian ${\cal H}$ determines the
energy spectrum of electrons in graphene under SMF-I. The system has
a periodicity of $2\pi$ as a function of $\delta$ due to gauge
invariance, so we restrict $\delta$ to range from $0$ to $2\pi$. The
solution to the DPs can be easily obtained: the original DPs located
at $k_x=0$, $\cos(\sqrt{3}k_y)=\frac{1}{2}-\cos\frac{\delta}{2}$,
for $0<\delta<4\pi/3$, and the newly generated DPs located at
$k_x=\pm\pi/3$,
$\cos(\sqrt{3}k_y)=-\frac{1}{2}-\cos\frac{\delta}{2}$, for
$2\pi/3<\delta<2\pi$.

\begin{figure}
\scalebox{0.6}[0.6]{\includegraphics[147,381][496,694]{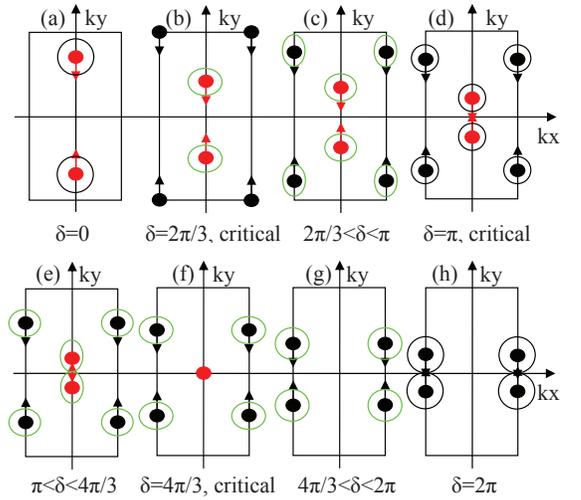}}
\caption{\label{fig2}(Color online) Schematic evolution of DPs in
MBZ with increasing staggered flux $\delta$ under SMF-I.
``{\color{red}$\bullet$}'' and ``$\bullet$'' represent the original
DPs in pristine graphene and the induced (additional) DPs
respectively, while the arrows represent their moving directions.
The black circles denote isotropic Dirac cones whereas the green
(grey) ellipses denote anisotropic Dirac cones. The coordinates of
the four corners of MBZ are ($\pm\pi/3, \pm\pi/\sqrt{3}$).}
\end{figure}

An overall picture of the evolution of DPs in magnetic Brillouin
zone(MBZ) under SMF-I is shown in Fig.~\ref{fig2}. When $\delta=0$,
the original pair of DPs (red filled circles) are located at
($0,\pm2\pi/3\sqrt{3}$) [Fig.~\ref{fig2}(a)]. As $\delta$ increases,
the two DPs move against each other along the $k_y$ direction, and
eventually they reach the center of MBZ at $\delta=4\pi/3$
[Fig.~\ref{fig2}(f)] and thereafter disappear [Fig.~\ref{fig2}(g)].
On the other hand, right at $\delta=2\pi/3$, one additional pair of
DPs are induced simultaneously at the four corners of MBZ (see the
black filled circles in the figure). Note that only two of the four
induced DPs are inequivalent and so only one pair contributes to the
system, since the four induced DPs are all located at the boundary
of MBZ. With increasing $\delta$ the two newly induced DPs move
along the lines $k_x=\pm\pi/3$ towards the edge center respectively
[Fig.~\ref{fig2}(c)-(h)]. When $\delta=2\pi$, the two DPs arrive at
($\pm\pi/3, \pm2\pi/3\sqrt{3}$). In this case, electrons hopping
along the arrows shown in Fig.~\ref{fig1}(a) will suffer an
additional phase $\pi/2$, which is a pure gauge. Thus the system
corresponding to this value of $\delta$ [Fig.~\ref{fig2}(h)] is
actually physical equivalent to that of Fig.~\ref{fig2}(a), because
they differ only by a gauge transformation.

\begin{figure}
\scalebox{1.2}[1.1]{\includegraphics[17,17][186,241]{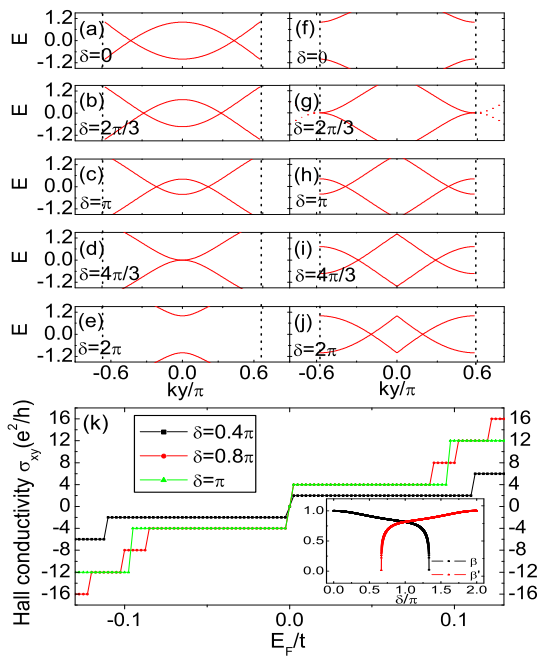}}
\caption{\label{fig3}(Color online) (a)-(e) Electron energy near the
DPs under SMF-I versus $k_y$ with $k_x=0$ for various values of
staggered flux $\delta$. The dashed line represents the boundary of
MBZ. (f)-(j)The same as (a)-(e) but with $k_x=\pi/3$. (k) Hall
conductivity $\sigma_{xy}$ under SMF-I, with $\phi=2\pi/768$ for
several values of $\delta$. Inset: The two renormalized factors
$\beta$ ($\beta'$) as functions of staggered flux $\delta$.}
\end{figure}

It is shown in Fig.~\ref{fig2} that DPs not only annihilate in pairs
but also emerge in pairs.\cite{Brey2009} This is interpreted by the
fact that the two DPs in each pair are connected to each other by
the time reversal symmetry which is still preserved by the SMF. Now
we lay out our numerical results to support these findings. We show
for different $\delta$ the energy dispersion near zero energy along
the two lines $k_x=0$ [Fig.~\ref{fig3}(a)-(e)] and $k_x=\pm\pi/3$
[Fig.~\ref{fig3}(f)-(j)] where DPs reside. Note that $\delta=2\pi/3$
and $4\pi/3$ are two critical values at which the pair of induced
DPs emerge and the original DPs completely superpose each other,
respectively. Apart from the two critical values, as $\delta$
increases from $0$ to $2\pi$, the number of DPs changes from one
pair to two pairs and then back to one pair. This interesting
evolution of DPs will dramatically affect the degeneracy of the LLs
which can be reflected by the Hall conductivity.

The Hall conductivity can be calculated directly through the
standard Kubo formula\cite{Thouless1982} by numerical
diagonalization of the Hamiltonian (1). In Fig.~\ref{fig3}(k), the
resulting Hall conductivity $\sigma_{xy}$ near zero energy is
plotted as a function of the Fermi energy $E_F$. According to the
Hall plateaus steps in $\sigma_{xy}$, the system can be classified
into three types. For $0<\delta<2\pi/3$ or $4\pi/3<\delta<2\pi$,
with spin degeneracy taken into account, $\sigma_{xy}$ has a step of
size $4e^2/h$, which is the same as that of pristine graphene, shown
in Fig.~\ref{fig3}(k) for $\delta=0.4\pi$. For
$2\pi/3<\delta<4\pi/3$, $\sigma_{xy}$ has a step of size $8e^2/h$
across zero energy (neutral filling) whereas a step of size $4e^2/h$
in the other energy range, which can be seen in Fig.~\ref{fig3}(k)
for $\delta=0.8\pi$. Remarkably, right at $\delta=\pi$, numerical
results of Hall conductivity show that all the steps have the same
size of $8e^2/h$. We interpret these phenomena as follows.

The isotropic Dirac cones become anisotropic under the influence of
a modulated magnetic field (see Fig.~\ref{fig2}). The chiral
fermions around an anisotropic Dirac cone can be physically
described by the anisotropic pseudospin Hamiltonian,
\begin{eqnarray}
\mathscr{H}=v_F\left(
\begin{array}{cc}
0&\hat{p}_-\\
\hat{p}_+&0
\end{array}\right)
\end{eqnarray}
where $\hat{p}_\pm=a\hat{p}_x\pm ib\hat{p}_y$, $v_F=3ta_0/2\hbar$ is
the Fermi velocity, and the two dimensionless coefficients $a$ and
$b$ measure the degree of anisotropy of the cone. In the presence of
a uniform magnetic field $B$ this anisotropy gives rise to a
renormalized LLs $E_n=\pm\beta\hbar v_F\sqrt{|n|}/l_B$ with
$\beta=\sqrt{ab}$ a dimensionless renormalization factor, and
$l_B=\sqrt{\phi_0/4\pi B}$ the magnetic length. With spin degeneracy
taken into account, it is found that for $0<\delta<4\pi/3$, the
4-fold degenerate LL spectrum near the original Dirac cones have a
$\beta$ value given by
$\beta=\beta(\delta)=\{\frac{2}{\sqrt{3}}\frac{1}{1+\cos(\delta/2)}(\frac{1+2\cos(\delta/2)}{3-2\cos(\delta/2)})^{\frac{1}{2}}\}^{\frac{1}{2}}$
(note when $\delta=0$, $\beta=1$), while for $2\pi/3<\delta<2\pi$,
the 4-fold degenerate LL spectrum near the induced Dirac cones have
another different $\beta'$ value given by
$\beta'=\beta(2\pi-\delta)$. For a general $\delta$ between $2\pi/3$
and $4\pi/3$, the LLs for the two branches are not degenerate except
the zeroth LL, which is exactly 8-fold degenerate. The zeroth LL is
independent of the external uniform magnetic field so its 8-fold
degeneracy cannot be removed, leading to a $8e^2/h$ Hall
conductivity step at the zeroth LL and a $4e^2/h$ step at other LLs.
However, when $\delta=\pi$, the two factors are equal to each other,
i.e., $\beta=\beta'$, all LLs for these cones (which are isotropic
now) overlap and so are exactly 8-fold degenerate. Therefore at
$\delta=\pi$, the Hall conductivity can be expressed as
$\sigma_{xy}=8(N+1/2)e^2/h$, with $N$ LL index. We remark that
actually, within the range of $\delta$ where the two pairs of Dirac
cones coexist, there should exist a series of critical values of
$\delta$ given by $\beta^2/\beta'^2=p/q$, where $p$ and $q$ are two
coprime integers. At these critical values, besides the zeroth LL,
the $mq$th LL (with $m=0,1,2,\ldots$ ) in the original pair of cones
are exactly degenerate with the $mp$th LL in the induced pair of
cones, giving rise to a $8e^2/h$ Hall conductivity step at these
energies. Another critical value of $\delta$ is at $\delta=4\pi/3$,
where the zero point at $k_x=k_y=0$ is not a DP, but rather a
semi-DP. Around the semi-DP, energy dispersion is found to be linear
along $k_x$, but parabolic along $k_y$. This peculiar feature can be
compared with the electronic structure in VO$_2$-TiO$_2$
nanoheterostructures.\cite{Banerjee2009}

\begin{figure}
\scalebox{0.6}[0.6]{\includegraphics[135,518][473,731]{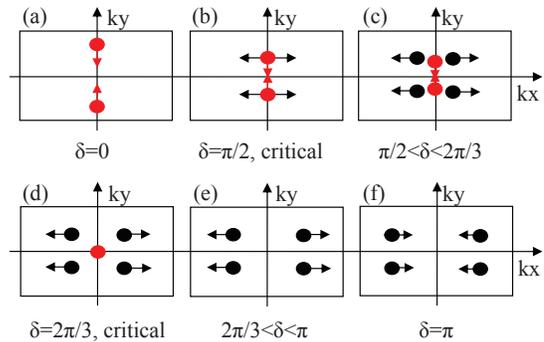}}
\caption{\label{fig4}(Color online) Schematic evolution of DPs in
MBZ with increasing staggered flux $\delta$ under SMF-II. All
symbols used here have the same meanings as that in Fig.~\ref{fig2}.
The coordinates of the four corners of MBZ are ($\pm\pi/3,
\pm\pi/2\sqrt{3}$).}
\end{figure}

Now we turn to explore the second type of the SMF shown in
Fig.~\ref{fig1}(b). Fig.~\ref{fig4} shows the schematic picture of
the evolution of DPs in the corresponding MBZ. Like that of SMF-I,
at the beginning of varying $\delta$, the two DPs of pristine
graphene are located at ($0, \pm\pi/3\sqrt{3}$), and then they move
towards the origin along $k_y$ direction. When $\delta=\pi/2$, each
DP of the pair changes into three DPs at k=($0,\pm\pi/6\sqrt{3}$)
[Fig.~\ref{fig4}(b)], but they completely superpose each other and
can not be distinguished there. After that the original DPs go on
moving along $k_y$ direction and eventually arrive at the origin at
$\delta=2\pi/3$ and then vanish, whereas the other two pairs of
induced DPs move along the $k_y=\pm\pi/6\sqrt{3}$ direction until
$\delta=\pi$ they reach the points ($\pm\pi/6,\pm\pi/6\sqrt{3}$)
[Fig.~\ref{fig4}(f)], and then backtrack. The evolution of DPs from
$\delta=\pi$ to $2\pi$ is just the reverse of the above process.

Compared with the SMF-I, in a period of $\delta$, there are four
critical values ($\delta=\pi/2, 2\pi/3, 4\pi/3$, and $3\pi/2$), at
which the DPs emerge or vanish. All the induced DPs under SMF-I move
parallel to the $k_y$ axis while that under SMF-II move parallel to
the $k_x$ axis. This should be associated with the configuration of
the SMF, where the induced DPs incline to move towards the periodic
direction of the SMF.

\begin{figure}
\scalebox{1.2}[1.1]{\includegraphics[17,17][186,239]{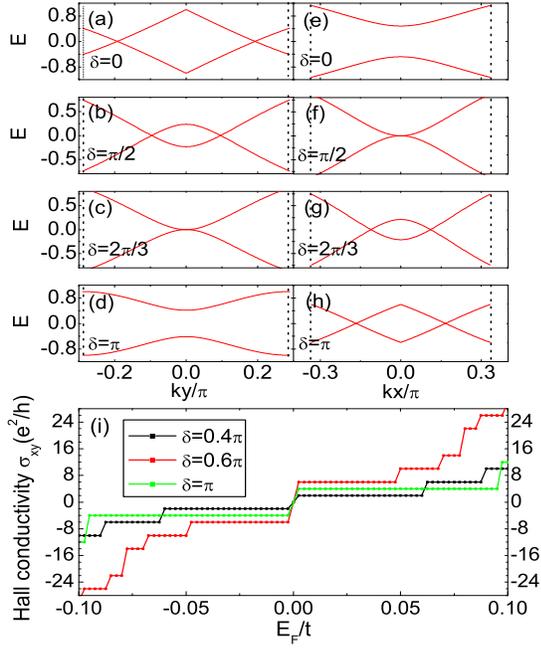}}
\caption{\label{fig5}(Color online) (a)-(d) Electron energy near the
DPs under SMF-II, as a function of $k_y$ with $k_x=0$ for various
values of staggered flux $\delta$. The dashed lines represent the
boundary of MBZ. (e)-(h)The same as (a)-(d) but as a function of
$k_x$ with $k_y=\pi/6\sqrt{3}$, instead. (i) Hall conductivity
$\sigma_{xy}$ under SMF-II with $\phi=2\pi/768$ for several values
of $\delta$.}
\end{figure}

The electronic energy spectrum near the zero energy are shown in
Fig.~\ref{fig5}(a)-(h). What is significant is that as $\delta$
increasing from $0$ to $\pi$, the number of DPs changes from one
pair to three pairs and then to two pairs. So, the DPs indeed emerge
and annihilate in pairs. In Fig.~\ref{fig5}(i), the Hall
conductivity $\sigma_{xy}$ is plotted as a function of the Fermi
energy $E_F$. For $0<\delta<\pi/2$ or $3\pi/2<\delta<2\pi$, the Hall
conductivity can be expressed as $\sigma_{xy}=4(N+1/2)e^2/h$, which
is the same as that of pristine graphene. This means that the
Dirac-cone topology is preserved within this range without new
induced DPs. For $\pi/2<\delta<2\pi/3$ or $4\pi/3<\delta<3\pi/2$,
$\sigma_{xy}$ has a $12e^2/h$ step across the zeroth LL and $4e^2/h$
or $8e^2/h$ step at the other LLs, implying the zeroth LL is 12-fold
degenerate. Interestingly, for $2\pi/3<\delta<4\pi/3$, all LLs are
8-fold degenerate and the Hall conductivity can be expressed as
$\sigma_{xy}=8(N+1/2)e^2/h$. This expression with quantized values
of half-integer multiples of $8e^2/h$ is robust and is interpreted
by the fact that the original pair of DPs has disappeared and the
induced two pairs of DPs are located symmetrically in MBZ giving
rise to exact 8-fold degeneracy of their corresponding LLs.

\begin{table}
\caption{\label{tab:table1}The evolution properties of DPs in MBZ at
$0<\delta<2\pi/L_x$ for various $L_x$'s ($L_x$ is from $2$ to
teens).}
\begin{ruledtabular}
\begin{tabular}{ccccc}
SMF $\delta$&$L_x$&$n_o$\footnote{$n_o$ represents the number of
pairs of DPs in MBZ with $k_x=0$.}(pairs)&$n_i$\footnote{$n_i$
represents the number of pairs of DPs in MBZ with
$k_x=\pm\pi/3Lx$.}(pairs)&$n_t$\footnote{$n_t$ ($=n_o+n_i$)
represents the total number of pairs of DPs in MBZ.}(pairs)\\
\hline
$0\leq\delta<\delta_c$\footnote{$\delta_c$ is a critical value where
the number of DPs at the line $k_x=0$ changes to zero. For
$L_x=2,3,\cdots,10$, $\delta_c$ is approximately
equal to $0.835\pi$, $0.595\pi$, $0.465\pi$, $0.375\pi$,$0.315\pi$, $0.275\pi$, $0.245\pi$, $0.217\pi$, $0.196\pi$, respectively.}&2&1&0,2&1,3\\
&3&1&0,2,3&1,3,4\\
&4&1&0,2,4&1,3,5\\
&5&1&0,2,4,5&1,3,5,6\\
&6&1&0,2,4,6&1,3,5,7\\
&7&1&0,2,4,6,7&1,3,5,7,8\\
&8&1&0,2,4,6,8&1,3,5,7,9\\
&9&1&0,2,4,6,8,9&1,3,5,7,9,10\\
&10&1&0,2,4,6,8,10&1,3,5,7,9,11\\
&$L$\footnote{$L\geq 2$.
For $L$ is an odd number, $n_i=0,2,4,\cdots,L-1,L$; for $L$ is an even number, $n_i=0,2,4,\cdots,L$.}&1&$0,2,\cdots,L$&$1,3,\cdots,L+1$\\
$\delta_c<\delta\leq 2\pi/L$&2&0&2&2\\
&3&0&3&3\\
&4&0&4&4\\
&5&0&5&5\\
&6&0&6&6\\
&7&0&7&7\\
&8&0&8&8\\
&9&0&9&9\\
&10&0&10&10\\
&$L$&0&$L$&$L$\\
\end{tabular}
\end{ruledtabular}
\end{table}

\begin{figure}
\scalebox{1.3}[1.2]{\includegraphics[17,16][174,196]{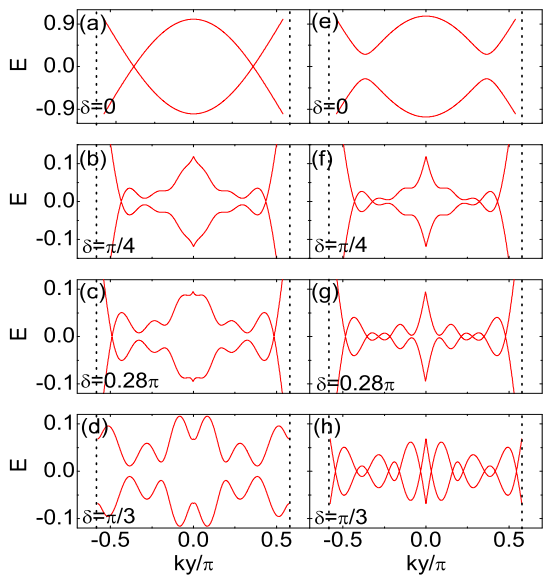}}
\caption{\label{fig6}(Color online) The case with $L_x=6$. Electron
energy near the DPs under a long-period SMF shown in Fig.~\ref{fig1}
(c) , as a function of $k_y$ with (a)-(d) $k_x=0$ and (e)-(f)
$k_x=\pi/18$ for various values of staggered flux $\delta$. The
figures in (b),(f) and (c),(g) have been shifted to the left by
$0.144\pi$ and $0.093\pi$, respectively. The total number of pairs
of DPs is 1,3,5,6, respectively for the four $\delta$ values.}
\end{figure}

Thus far we have discussed two simplest SMFs where the magnetic flux
alternates along the armchair chains or zigzag chains, respectively.
Now we generalize our theory to the cases of SMF with long spatial
period, shown in Fig.~\ref{fig1}(c). Taking SMF-I for example, we
make the magnetic flux alternate every $L_x$ zigzag chains.
Numerical analysis shows that, as $\delta$ increases from $0$ to
$2\pi/L_x$,\footnote{Here we only consider the weak magnetic flux
case $\delta<2\pi/L_x$ for two reasons. One is the weak magnetic
flux can be easily achieved in experiment; the other is that for
strong magnetic flux, it is hard to find a good rule to describe the
evolution of DPs.} the number of pairs of original DPs $n_o$ (at
$k_x=0$) decreases from one to zero, while that of the induced DPs
$n_i$ (exactly located at the boundary of MBZ $k_x=\pm \pi/3L_x$)
increases from zero to $L_x$ gradually obeying a sequence of
$0,2,4,\cdots, L_x$. Accordingly, the total number of pairs of DPs
is $n_t=1,3,5,\cdots,L_x+1,L_x$. For the detail, see Table I. In
Fig.~\ref{fig6}, we take $L_x=6$ for example. When
$0<\delta<2\pi/L_x$, the number of pairs of the induced DPs shows a
sequence of $n_i=0, 2, 4, 6$ [Fig.~\ref{fig6}(e)-(h)], while that of
the original pair is always $n_o=1$ or $0$ [Fig.~\ref{fig6}(a)-(d)].
Once again, the DPs emerge and annihilate in pairs.

After application of such a long-period SMF, each pair of Dirac
cones has generally different anisotropy, i.e., has different
renormalization factor $\beta$ in their corresponding LL spectrum,
so all LLs except zeroth LL are still 4-fold degenerate. However,
the zeroth LL is exact $4n_t$-fold degenerate, leading to a step of
size $4n_te^2/h$ in $\sigma_{xy}$ at the zeroth LL,\footnote{The
Hall plateaus are a little different from that discussed in
Ref.\onlinecite{Park2009}. In particular, when $L_x$ is an even
number, there is a $4L_xe^2/h$ Hall conductivity step.} as $\delta$
increasing from $0$ to $2\pi/L_x$. Therefore, under a modulated
orbital magnetic field, the property of neutral graphene is actually
governed by the number of pairs of DPs. This is a significant
signature and can be detected by Hall measurements. So long as $L_x$
is no more than the magnetic length of the systems, the physics
contained in them is similar. But for graphene system with the
magnetic length much less than $L_x$, the DPs will become more and
more dense and will finally be merged into the zeroth LL of
graphene.\cite{Xu} This deserve further study and will be discussed
elsewhere.

In summary, we have investigated the electronic structure in neutral
graphene under periodically modulated magnetic fields. It is found
the modulated magnetic field can induce additional DPs in graphene
and the evolution of these DPs can be manipulated by the magnitude
and period of the field. These induced DPs add additional degeneracy
to the LLs, especially a $4n_t$-fold degeneracy at the zeroth LL,
leading to an unusual integer QHE near neutral filling. These
phenomena are expected to be observed by Hall measurements.

\begin{acknowledgments}
L. X. thanks Y. Zhou and Y. Zhao for useful discussion. This work
was supported by NSFC Projects 10504009, 10874073 and 973 Projects
2006CB921802, 2006CB601002.
\end{acknowledgments}


\end{document}